\documentstyle[12pt,epsfig]{article}
\begin{document}
\begin{titlepage}
\begin{center}

\vspace*{2cm}
{ \Large \bf A remark on the low mass dilepton yield\\

in heavy ion collisions\footnote{Presented at the XXXVI ISMD, Paraty, Brasil, September 2-8, 2006,

to be published in the conference proceedings in Brazilian Journal of Physics} }

\vspace{2cm}

\begin {author}
\Large K. Fia{\l}kowski\footnote{e-mail address:
uffialko@th.if.uj.edu.pl}
\end{author}

\vspace{1cm}
{\sl M. Smoluchowski Institute of Physics, Jagellonian University,\\
30-059 Krak{\'o}w, ul.Reymonta 4, Poland}

\vspace{2cm}

\begin{abstract}
The recent data on the enhancement of the low mass dilepton
yield in heavy ion collisions are interpreted as an effect of the "prolonged
life" of resonances in the hadron gas phase.  The value of the enhancement
factor gives an upper limit for the duration time of this phase.

\end{abstract}
\end{center}
\vspace{1cm}

PACS: 25.75.-q\\

{\sl Keywords:}  Heavy ion collisions, dileptons  \\

\vspace{0.5cm}

\end{titlepage}

\section{Introduction}

\par In the last decade a lot of attention has been paid to the spectra of dileptons produced in heavy ion collisions.  In
particular, CERES data of 1995-1996 \cite{CERES1} were extensively analyzed and discussed both by the CERES collaboration
(which released only recently the definite version of their results \cite{CERES2}) and by many theorists \cite{TH}. Whereas 
the spectra from $pBe$ and $pAu$ collisions agree satisfactorily 
with a "decay cocktail" $R_c (m)$ ($R$ 
calculated from the known yield of various hadrons and their decay processes, 
dominated by $\rho$ in the mass range 
$0.6-0.9~ GeV/c^2$), already the $SAu$ collisions show a clear excess of dileptons 
with masses in the range $0.2-0.8~ 
GeV/c^2$, and no clear peak at the $\rho$ mass \cite{CERES2}. More
recently, the dimuon spectra from the In-In ion collisions were presented \cite{NA60}
from an experiment which collected thousand times as many events. However, these data are 
not normalized to the charged hadron spectra. The excess of dileptons was established 
assuming no excess for the $\omega$ and $\phi$ peaks and the standard ratio $\rho/\omega\simeq 1.2$
for the "cocktail" contribution. In this
note we suggest to use the dilepton results for a very simple test of basic assumptions concerning the space-time
development of particle production in heavy ion collisions.
\par  Let us remind first that many hadron-nucleus data are 
reasonably described if the process is treated in the first approximation as a
superposition of nucleon-nucleon interactions (more precisely, an idea of "wounded nucleons" is used, where the number
of interacting - "wounded" - nucleons, and not the number of interactions, determines the multiplicity) \cite{WN,WN2}.  
This idea was also implemented successfully in the string-like MC generators \cite{LUND}. On the other hand, nucleus-nucleus 
(heavy ion) collisions are known to produce more particles, than the "wounded nucleon model" predicts, and many "collective 
effects" contradicting the superposition picture were found to exist \cite{Roland}.  There are many proposals to cure this 
problem ("wounded quarks" \cite{PH2}, string interactions leading to "color ropes" \cite{CR} etc.), but no satisfactory 
global description of data in this framework was presented.  
\par Therefore the different description is used, where the collision is supposed to produce (after some preliminary stages) a
"hadron gas" expanding and cooling thermodynamically (in a chemical equilibrium?)  until the "freeze-out" stage, from which
hadrons flow freely to the detectors (obviously subject to decays on their way) \cite{HAG}.  Both the transverse spectra and
the relative hadron multiplicities were successfully described in this picture \cite{BF} as well as many other data
\cite{HR}. However, this fact cannot be
regarded as disqualifying definitely a "superposition" picture, as it is possible to imitate "thermal spectra" of
transverse momenta by the string decays, when the string tension fluctuations are taken into account \cite{BIALAS} (in fact,
similar spectra are seen for elementary collisions \cite{Becat}).  Also the "chemical composition" of hadrons may be
explained similarly. 
\par In the last decade the majority of physicists accepted the suggestion that the collision leads to a "quark-gluon 
plasma formation" \cite{QGP} (possibly preceded by a
"color glass condensate" state \cite{McL}, whose interaction explains some features of the plasma state).  The evolution of
this plasma, ruled by the equations of QCD, leads eventually to the hadronization, possibly followed by a "hadron gas"
phase before the "freeze-out".  However, some characteristics of plasma survive in measured hadron quantities \cite{QGPsig}.
\par Dileptons belong to the small class of signals which may collect the contributions from all the stages of the process,
since the lack of strong interactions allows them to escape both from "quark-gluon plasma" and from "hadron gas", and,
obviously, they may result from the electromagnetic decays of hadrons on their way to detectors.  In the case of neutral
pions, for which no competition from strong decays exists, this last process is certainly a dominant component, and the
"lowest mass" dilepton spectrum is simply proportional to the number of final pions (and thus rather independent of the
adopted assumptions on production process).  This is, however, not the case for heavier mesons, for which the branching
ratios for the electromagnetic decays are small.  In particular, the dileptons from $\rho$ decay were investigated and
their yield was found to depend crucially on the process.

\section{Data and their interpretation}

\par The data are usually shown as the ratio of the dilepton
yield (as a function of dilepton mass) to the charged hadron yield for the same range of rapidity and transverse momentum

$$R(m)=\frac{\int{dyd^2p_t \frac{dN_{ee}}{dyd^2p_tdm}}}{\int{dyd^2p_t \frac{dN_{ch}}{dyd^2p_t}}}.$$

\par As already noted, the $SAu$ CERES data show a clear excess of dileptons 
with masses in the range $0.2-0.8~ 
GeV/c^2$, and no clear peak at the $\rho$ mass \cite{CERES2}. This contradicts 
the "superposition" scenario, unless very serious corrections are introduced.
\par The $PbAu$ data for semi-central events ($28\%$) \cite{CERES2} reinforce this conclusion. Below the mass 
of $0.2~ GeV/c$ (where the $\pi^0$ decay dominates)  the data agree well with the "decay cocktail"
and the ratio of the integrated yields is compatible with one 

$$\overline R=\frac{\int_0^{0.2} R^{PbAu}(m)dm}{\int_0^{0.2} R^c(m)dm}=0.92 \pm 0.17$$ with the error dominated by systematical
effects.  
\par However, for the mass range between $0.2$ and $0.9~GeV/c^2$ the ratio of integrated yields is significantly
bigger than $1$

$$\overline R=\frac{\int_{0.2}^{0.9} R^{PbAu}(m)dm}{\int_{0.2}^{0.9} R^c(m)dm}\simeq 2.5$$
(with combined systematic and decay uncertainty of 
the order of $1$). The excess seems to grow with centrality. Again, no clear 
peak at the $\rho$ mass is visible.  
The data for transverse pair momenta below $0.5~ GeV/c$ are obviously similar to 
the spectra integrated over full $p_t$ range (which they dominate); for higher $p_t$ the shape is more 
similar to the "cocktail" curve, but the enhancement factor for the yield integrated over 
all the $0.3-0.9~ GeV/c^2$ mass range does not depend visibly on $p_t$.
\par These results are interpreted by the CERES group \cite{CERES2} as an 
evidence for two effects. "Medium-modification" of the $\rho$ 
spectral function is responsible for the disappearance of the peak at $\rho$ mass.
The enhancement of the global dilepton yield is 
interpreted as due to the binary $\pi \pi$ (or 
$q\overline q$) annihilation processes (growing faster than linearly with the pion density).
\par We agree that the strong change of shape of the mass spectrum suggests 
the need of some modification of the $\rho$ 
contribution (mass decrease and/or width increase). 
However, we would like to point out that there is 
an obvious reason for the enhancement of the global yield other than the 
"binary annihilation processes".
\par This reason is the very existence of "chemical equilibrium" phase, 
where the ratio of $\rho$/$\pi$ is governed by the 
temperature and thus strong decays of $\rho$ are effectively balanced 
by the formation of $\rho$ in the $\pi\pi$ 
collisions. If for the freely decaying $\rho$ the branching ratio 
for decay into dielectrons is about $5\cdot10^{-5}$, the 
number of electromagnetic decays from the $\rho$ "forced to live 
longer" will be obviously higher. In the first 
approximation, one expects in this mass range the enhancement factor
 $$\overline R=t/\tau,$$ 
where $t$ is the lifetime of "hadronic gas" phase, and $\tau$ the free $\rho$
lifetime ($\tau \simeq 1 fm/c$). In fact, an effective Lorentz factor should be included here, 
but for the central rapidity bin its value is not much different from one.

\par The remark equivalent to the relation proposed above was made recently by Renk and Ruppert \cite{Renk},
who argue that the dilepton data may be successfully described and used to discriminate between various models
of "in-medium modifications of vector mesons". However, we are convinced
that simple relation may be useful without invoking any specific model.

\par Obviously, our reasoning is rather simplistic. The spectral function is modified strongly
in the hadron gas phase. If the $\rho$ mass 
is reduced  and/or its shape changed "in-medium", the branching ratio for electromagnetic decay may be 
also changed; the $\rho /\pi$ ratio should not be constant, but rather decrease with decreasing 
temperature during the evolution of "hadron gas" 
phase. However, we feel that the suggested value of $t \simeq 2.5~fm/c$ (much smaller 
than the global lifetime of the "fireball" measured 
by Bose-Einstein interference effects) is rather intriguing.
\par Let us stress here once more that we discuss only the dilepton yield integrated over rather wide
mass range, and not the shape. Thus the detailed changes of the spectral function are irrelevant for
our argument. We compare just the dilepton yield (proportional to the four dimensional volume) and the hadron 
yield (proportional to the three dimensional volume), thus estimating the life time of the hadron gas phase.
\par If the 
significant part of the observed dilepton spectra 
originates from the "binary annihilation processes", as suggested in \cite{CERES2},  
the value of $t$ should be even smaller, since 
obviously
$$R^{AA}(m)=R^{AA}_{hg}(m)+R^{AA}_{ann}(m)$$
and thus
$$\frac{t}{\tau}=\frac{R^{AA}_{hg}}{R^c}<\frac{R^{AA}}{R^c} \simeq 2.5.$$

\par The moderate value of the "enhancement factor" makes the statement 
about the rejection of "superposition scenario" less 
categorical. One may imagine that some coalescence of "color strings" 
into "color ropes" characterized by larger transverse 
dimensions, higher string tension and its fluctuations may lead to the 
states quite similar to the "hadron gas phase" (and/or "quark gluon plasma phase"). Their 
lifetime may be easily of the order of few $fm/c$, but still much shorter 
than the global time of the hadroproduction process.
\par In the high statistics NA60 experiment \cite{NA60}
 a similar value of the
enhancement factor (increasing with centrality from $\simeq 1.5$ to $4$) was found.
The new data suggest, however, that the $\rho$ peak is broadened rather than moved to lower
mass values in comparison to the free $\rho$ decay.
\par The NA60 analysis agrees qualitatively with the naive expectations 
from the "hadron gas enhancement effect".
If the duration time of hadron gas phase is of the order of few fm/c, the enhancement of the
$\omega$ contribution is very mild, and for the $\phi$, $\eta$ or $\eta '$ completely negligible,
as the lifetimes of these particles are much longer (however, the annihilation processes 
contribute significantly for higher masses increasing the experimental value
of $\overline R$). Thus the NA60 data seem to support our 
suggestions that the main reason of the enhancement 
of the dilepton yield in the $\rho$ mass region is the "extended lifetime" of $\rho$.

\section{Comments and conclusions}
\par Our suggestion tells nothing about the "quark-gluon plasma phase", 
if it does not contribute dominantly to 
the low mass dilepton production. However, the small lifetime of "hadron gas" phase 
suggests that the plasma phase (and possibly other state) is likely to precede it.
\par An intriguing difference between the spectra for dilepton transverse momenta 
below- and above $0.5~ GeV/c$ (qualitatively explained within the formalism of the 
"in-medium modification" \cite{CERES2}) does not 
extend, as noted above, to the integrated yield. The "enhancement factors" for the full mass range 
are nearly the same in both cases. Thus no 
significant difference between the freeze-out times for two ranges of $p_t$ is suggested.
 The NA60 
data, on the other hand, imply a stronger $p_t$ dependence of the enhancement factor. 
\par Summarizing, we suggest that main reason for the enhancement of the ratio of the number of 
the low mass dilepton pairs to the charged 
hadron density for heavy ion collisions is just the existence of 
the "hadron gas phase" in which there is a 
chemical equilibrium between pions and heavier mesons. However, the observed 
moderate value of the "enhancement factor"  
leads to the conclusion that this phase is rather short-lived. A detailed 
quantitative analysis of this problem may be very 
helpful for the proper description of the space-time development of the 
hadroproduction processes in heavy ion collisions.

\noindent{\bf Acknowledgements}

\par I would like to thank Romek Wit, Andrzej Kotanski and in particular 
Andrzej Bia{\l}as for reading the manuscript and helpful remarks. I am grateful to Torsten Renk
and Sanja Damjanovic for bringing to my attention refs.\cite{Renk} and \cite{NA60}, respectively. 
Critical remarks of Ralf Rapp are gratefully acknowledged.
This work has been partly supported by the Polish Ministry of Education and Science
grant 1P03B 045 29 (2005-2008).

\end{document}